\documentclass[preprint,12pt]{JHEP3}
\usepackage{axodraw}

\newcommand\lsim{\mathrel{\rlap{\lower4pt\hbox{\hskip1pt$\sim$}}
    \raise1pt\hbox{$<$}}}
\newcommand\gsim{\mathrel{\rlap{\lower4pt\hbox{\hskip1pt$\sim$}}
    \raise1pt\hbox{$>$}}}
\def\bea{\begin{eqnarray}}
\def\eea{\end{eqnarray}}
\def\ba{\begin{array}}
\def\ea{\end{array}}
\def\bc{\begin{center}}
\def\ec{\end{center}}
\def\nn{\nonumber}

\def\f{\frac}

\def\e{\epsilon}

\def\l{\lambda}
\def\r{\rho}
\def\f#1#2{\frac{#1}{#2}}

 \preprint{SNUTP 05-07}
\title{Hiding An Extra Dimension}
\author{Hyung Do Kim\\
School of Physics and Center for Theoretical Physics,\\
Seoul National University, \\
Seoul, 151-747, Korea\\
E-mail: \email{hdkim@phya.snu.ac.kr}}
\abstract{We propose a new geometry and/or topology of a single extra
dimension whose Kaluza-Klein excitations do appear at much higher
scale than the inverse of the length/volume.
For a single extra dimenion with volume $N\pi\rho$ which is
made of $N$ intervals with size $\pi \rho$ attached at one point,
Kaluza-Klein excitations can appear at $1/\rho$ rather than $1/N\rho$
which can hide the signal of the extra dimenion sufficiently
for large $N$.
The geometry considered here can be thought of 
a world volume theory of self intersecting branes or an effective description
of complicated higher dimensional geometry such as Calabi-Yau with genus
or multi-throat configurations.
This opens a wide new domain of possible compactifications
which deserves a serious investigation.
}

\begin{document}
%%%%%%%%%%%%%%%%%%%%%%%%%%%%%%%%%%%%%%%%%%%%%%%%%%%%%%%%

%%%%%%%%%%%%%%%%%%%%%%%%%%%%%%%%%%%%%%%%%%%%%%%%%%%%%%%%
\section{Introduction}
%%%%%%%%%%%%%%%%%%%%%%%%%%%%%%%%%%%%%%%%%%%%%%%%%%%%%%%%

Unification of gauge and gravitational interactions is one
of the most important paradigm in particle physics
and it has guided theoretical physics when the experiments
did not follow theory. Three gauge couplings are believed to be
unified at very high energy so called grand unification scale
(GUT scale). In the standard model it works within
10 to 20 percent errors
and in the minimal supersymmetric extensions of it,
the unification works a lot better (within a few percent errors).
Thus it seems to provide a strong hint for what is new physics
at TeV scale or higher.
In order to unify gauge interactions with gravity,
first we should understand why the electroweak scale is so lower
compared to the Planck scale at which gravitational interactions
become of order one similar strength to the gauge interactions.
Supersymmetry broken at TeV is regarded as the most popular solution
to this problem. 

However, we can address the question in a different way.
Why is gravity so weak? Effective gravitational interaction
at given energy scale is $E^2/M_{Planck}^2$ and is extremely tiny
compared to order one gauge interactions. 
This question brought entirely new solutions to the problem
of disparity between gravity and gauge interactions
in terms of extra dimenions.
Large extra dimension \cite{Arkani-Hamed:1998rs}\cite{Antoniadis:1998ig}
explains the weakness of gravity
in terms of large volume of extra dimenions only gravity feels.
Warped extra dimension (a slice of $AdS_5$) proposed by Randall and Sundrum
\cite{Randall:1999ee}
naturally provides TeV brane at which the natural scale is just TeV
due to an exponential warp factor along the extra dimension.
Graviton zero mode wave function is not flat in $AdS_5$
but is localized at Planck brane. Thus TeV brane matter feels only
the tail of graviton zero mode and weakness of gravity is naturally explained
even with a small (order one) size of the extra dimension.

Flat extra dimension with size smaller than 0.1mm is consistent
with the current experimental limit \cite{Adelberger:2003zx}
as long as gauge interactions
are confined on the brane and only gravity feels it.
Submillimeter extra dimensions make gravity be strong at TeV
if there are two extra dimensions which is just the limit
from precision gravity experiment.
Although it provides the most interesting possibility,
there comes a strong constraint from astrophysics/cosmology.
From the supernovae and neutron stars we would expect more gamma rays
from decays of massive Kaluza-Klein gravitons
whose mass is below the temperature of the supernovae core, 30 MeV.
This puts the most stringent bound on large extra dimensions
\cite{Hannestad:2001xi}.
Single extra dimension gives too light massive graviton
which is already inconsistent with the experimental fact
if we force the scale of quantum gravity at around TeV.
For two extra dimensions, the bound pushes the scale of quantum gravity
beyond 1000 TeV and we can not relate it to the weak scale any longer.
In this paper we suggest a setup in which the lightest Kaluza-Klein
graviton is heavy enough and can be consistent with the experimental bounds.
In this setup the N-fold degeneracy with sufficiently large N
provides a rapid change of the gravitational interactions
such that gravity can be of order one at TeV.

String theory is usually defined in 10/11 dimensions
and 6/7 extra dimensions should be curled up and be hidden
to be consistent with the fact that we live in 3+1 noncompact spacetime.
The most popular scenario assumes Calabi-Yau space as
the compactification manifold to yield 4D N=1 supersymmetry
\cite{Candelas:1985en}.
Recently compactification with various flux has been intensively studied
as it provides the stabilization of most string theory moduli
which otherwise would remain massless \cite{Klebanov:2000hb}
\cite{Giddings:2001yu}\cite{Kachru:2003aw}\cite{Kachru:2003sx}.
Flux compactification also generates throat geometry in Calabi-Yau
and the long throat physics is well described in terms of
effective 5 dimensional theory.
Full 10 dimensional physics appears only at very high energy scale
near the string scale and the low energy excitations are just
the Kaluza-Klein states of Randall-Sundrum like setup.
It is then natural to imagine that there would be many throats
in Calabi-Yau space and we can ask what the theory looks like
if Calabi-Yau has multi-throat geometry.
In this case we have a clear distinction between scales
of Kaluza-Klein excitations and light modes appear only at
around infrared(IR) branes.
There are many physical questions that can be addressed without
knowing full 10 dimensional spectrum.
Therefore it would be interesting to see what the spectrum will look like
for the multi-throat geometry.
The essential property of multi-throat geometry is kept
when we replace each throat by RS geometry which just include
single extra coordinate 
\cite{Dimopoulos:2001ui}\cite{Dimopoulos:2001qd}\cite{Cacciapaglia:2005pa}.
\footnote{Recent studies are in
\cite{Kofman:2005yz}\cite{Bruemmer:2005sh}.}
Then the bulk region corresponds to the ultraviolet (UV) brane.
As all the throats are connected to the bulk,
several IR branes are linked to the UV brane
through the slice of $AdS_5$.
This setup is exactly the one we will study here.

Once we have a situation where the extra coordinate is just one
but has a several branch starting from the UV brane,
we can generalize it to the flat space.
The junction of extra space is nothing to do with the curvature
of each $AdS_5$ and we can attach several different $AdS_5$ slice
with different curvatures at the same time.
Therefore, it is natural to imagine the flat limit of the same configuration.
At least we can define a consistent field theory on the flat limit
of the multi-throat effective theory and can study the theory on it.
How to get such a geometry from Calabi-Yau or other compactification
is an independent question and we will not address it here.
One obvious example is the torus with a genus one.
When one cycle wrapping the genus is much larger than the other cycle,
we can approximate the geometry as one dimensional ring at low energy scale.
The excitation associated to the other cycle will appear
only at very high energy scale and will be irrelevant to the physics
below the inverse scale of the other cycle.
We can find an effective 5 dimensional description
of multi-geni Calabi-Yau in a similar way.

In this paper we will analyze the spectrum of the fields
living in a single extra dimension discussed above.
After a brief discussion on how to get such an extra dimension,
we use deconstruction with a few sites for the analysis.
We also study the phenomenology with spectrum obtained by 
deconstruction technique.
Then we discuss the actual analysis in field theory.
Finally we conclude with a few remarks.

%%%%%%%%%%%%%%%%%%%%%%%%%%%%%%%%%%%%%%%%%%%%%%%%%%%%%%%%
\section{Brane intersection of its own}
%%%%%%%%%%%%%%%%%%%%%%%%%%%%%%%%%%%%%%%%%%%%%%%%%%%%%%%%

As long as gauge interactions are concerned, the best way 
to obtain the flat space limit of multi-throat geometry
is the brane intersection of its own.
We consider a setup in which a brane bends and finally intersects
by itself. The simplest possibility is to have figure eight(8). We can
continue the process such that many rings intersect at a single
point. Perhaps the most simplest one is to fold the ring such that
there would be an interval. The final setup would be the gathering
of many intervals with one common point. Suppose that the
individual interval has a finite length $\r$ and there are $N$ such
intervals. The total length is then $N\r$. Any gauge theory living
on this configuration would have a suppression $1/(N\r)$ in its 4D
gauge coupling. Now the question is the scale of Kaluza-Klein
excitations.

Thus we consider these configurations. To see the new feature
clearly, we take the deconstruction 
\cite{Arkani-Hamed:2001ca}\cite{Hill:2000mu}
as our analysis tool.

%%%%%%%%%%%%%%%%%%%%%%%%%%%%%%%%%%%%%%%%%%%%%%%%%%%%%%%%
\subsection{Deconstruction}
%%%%%%%%%%%%%%%%%%%%%%%%%%%%%%%%%%%%%%%%%%%%%%%%%%%%%%%%

If we do the analysis for the circle moose diagram, we would
obtain the eigenvalues \bea M^2_n & = & (\f{2}{a})^2 \sin^2
(\f{na}{2R}), \ \ \ \f{-N}{2} < n \le \f{N}{2} \nn \eea where $a =
\f{1}{g \langle \Phi \rangle}$ and $R=Na$. For $N \gg 1$ and $n \ll N$,
the expression is well approximated to be \bea M^2_n & = &
(\f{n}{R})^2. \nn \eea

%%%%%%%%%%%%%%%%%%%%%%%%%%%%%%%%%%%%%%%%%%%%%%%%%%%%%%%%
\subsubsection{N-Octopus}
%%%%%%%%%%%%%%%%%%%%%%%%%%%%%%%%%%%%%%%%%%%%%%%%%%%%%%%%

First of all, suppose there is a center point at which several
intervals are connected. We call it 'octopus' diagram although the
legs need not be eight. Let the legs be N. Each leg has one end
adjacent to the head of the octopus (the center). The boundary
condition would determine the eigen modes along the extra
dimension but it would be easier to see it from a simplified
deconstruction setup.

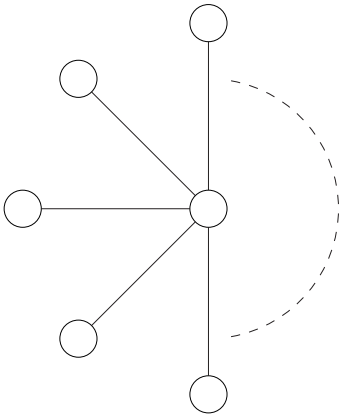
\SetScale{0.7}
\begin{center}
\begin{picture}(300,200)(100,50)
\Line(200,200)(200,300)
\Line(200,200)(200,100)
\Line(200,200)(100,200)
\Line(200,200)(130,270)
\Line(200,200)(130,130)
\GCirc(200,200){10}{1}
\GCirc(200,300){10}{1}
\GCirc(200,100){10}{1}
\GCirc(100,200){10}{1}
\GCirc(130,270){10}{1}
\GCirc(130,130){10}{1}
\DashCArc(200,200)(70,-80,80){5}
%\Text(150,80)[]{N-Octopus moose diagram}
\end{picture}
\end{center}

Let us consider a gauge theory on it. There is a gauge boson
$A_{\mu}^0$ which is at the head and each leg connects $A_{\mu}^0$
to $A_{\mu}^i$ where $i = 1, \cdots, N$. If the scalar fields
linking two sites get VEVs, the corresponding gauge bosons become
massive. The link field $\Phi_i$ is bi-fundamental under the gauge
group $G^0$ and $G^i$. The mass matrix for $N+1$ gauge bosons is \bea
M^2 & = & \f{1}{a^2} \left( \ba{cccccc}
1 & 0 & 0 & \cdots &  0 & -1 \\
0 & 1 & 0 & \cdots &  0 & -1 \\
0 & 0 & 1 & \cdots &  0 & -1 \\
\cdots & \cdots & \cdots & \cdots & \cdots & \cdots \\
0 & 0 & 0 & \cdots  & 1 & -1 \\
-1 & -1 & -1 & \cdots & -1 & N \ea \right) \eea where $a =
\f{1}{g\langle \Phi \rangle}$ and the $N+1$th column and row
correspond to $A_{\mu}^0$. There are $N+1$ eigenstates. The
characteristic equation can be easily derived for $\hat{M}^2 = a^2
M^2$. 
\bea \det (\hat{M}^2 - \lambda I ) = \l (1-\l)^{N-1} \{ \l -
(N+1)\} 
\eea 
There is a zero mode $\lambda = 0$ with the
eigenvector $v_0 = \f{1}{\sqrt{N+1}}(1,1,1,\cdots,1)$. The
lightest Kaluza-Klein states are degenerate. There are $N-1$
states with mass $(\f{1}{a})^2$. The eigenvectors should be
orthogonal to the zero mode and its $N+1$th component is zero.
Thus $v_1 = \f{1}{\sqrt{2}}(1,-1, 0, \cdots,0,0,0,\cdots, 0)$,
$v_2 = \f{1}{\sqrt{6}}(1,1,-2, \cdots,0,0,0,\cdots, 0)$ and $v_i =
\f{1}{\sqrt{i(i+1)}} (1,1,1,\cdots,1,-i,0,\cdots,0)$ where
$i=1,\cdots,N-1$. (The final one with $i=N$ is not linearly
independent if there are vectors from $i=1$ to $i=N-1$.) The last
one has the eigenvalue $\f{(N+1)}{a^2}$ and the eigenvector is
$v_{N} = \f{1}{\sqrt{N(N+1)}} (1,1,\cdots,-N)$.

The deconstruction for the octopus with N legs can be easily
generalized to include higher excitations of each leg by adding
more sites between the site $0$ and $i$. The octopus has two
distance scales. One is the size of each leg $\r$ which is just the
lattice size in the above example $\r=a$. The other is the total
volume of the extra dimension which is simply $N$ times $\r$.
($R=N\r$). As the total length is the longest one, you might guess
that the lowest excitation will appear at a scale $1/R$ but it
turns out that it appears only at $1/\r = N/R$. It is an
interesting example in which the volume suppression can be large
and at the same time the Kaluza-Klein excitations associated with
it can be very heavy.\footnote{With two or more extra dimensions,
distinct KK modes appear if we consider compact hyperbolic extra dimensions
\cite{Kaloper:2000jb}.}

%%%%%%%%%%%%%%%%%%%%%%%%%%%%%%%%%%%%%%%%%%%%%%%%%%%%%%%%
\subsubsection{Two Centers}
%%%%%%%%%%%%%%%%%%%%%%%%%%%%%%%%%%%%%%%%%%%%%%%%%%%%%%%%

Let us consider the second example in which there are two centers.

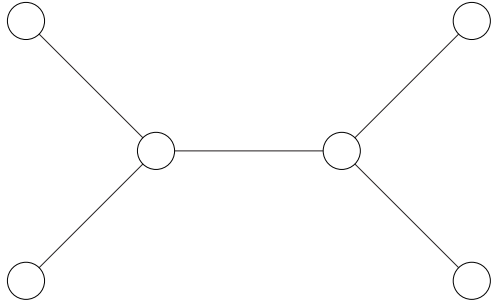
\SetScale{0.7}
\begin{center}
\begin{picture}(300,150)(0,0)
\Line(100,100)(200,100)
\Line(100,100)(30,170)
\Line(100,100)(30,30)
\Line(200,100)(270,170)
\Line(200,100)(270,30)
\GCirc(100,100){10}{1}
\GCirc(200,100){10}{1}
\GCirc(30,170){10}{1}
\GCirc(30,30){10}{1}
\GCirc(270,170){10}{1}
\GCirc(270,30){10}{1}
%\Text(100,10)[]{Two centers with two legs each}
\end{picture}
\end{center}

It is straightforward to generalize the setup.

\bea M^2 & = & \f{1}{a^2} \left( \ba{cccccc}
1 & 0 & -1 & 0 & 0 & 0 \\
0 & 1 & -1 & 0 & 0 & 0 \\
-1 & -1 & 3 & -1 & 0 & 0 \\
0 & 0 & -1 & 3 & -1 & -1 \\
0 & 0 & 0 & -1 & 1 & 0 \\
0 & 0 & 0 & -1 & 0 & 1 \ea \right)  \eea

We can list the eigenvalues and the eigenstates for $\hat{M}^2 =
a^2 M^2$ up to normalization.

\bea \l = 0 & (1,1,1,1,1,1) \nn \\
\l = \f{5-\sqrt{17}}{2} & (1,1, \f{\sqrt{17}-3}{2},
-(\f{\sqrt{17}-3}{2}), -1,-1) \nn \\
\l = 1 & (1,-1,0,0,0,0) \nn \\
\l = 1 & (0,0,0,0,1,-1) \nn \\
\l = 3 & (1,1,-2,-2,1,1) \nn \\
\l = \f{5 +\sqrt{17}}{2} & (1,1,-(\f{3+\sqrt{17}}{2}),
\f{3+\sqrt{17}}{2}, -1,-1) \nn \eea

%%%%%%%%%%%%%%%%%%%%%%%%%%%%%%%%%%%%%%%%%%%%%%%%%%%%%%%%
\subsubsection{Two Centers with 2N Legs}
%%%%%%%%%%%%%%%%%%%%%%%%%%%%%%%%%%%%%%%%%%%%%%%%%%%%%%%%

\SetScale{0.7}
\begin{center}
\begin{picture}(300,150)(0,0)
\Line(100,100)(200,100)
\Line(100,100)(30,170)
\Line(100,100)(15,150)
\Line(100,100)(30,30)
\Line(200,100)(270,170)
\Line(200,100)(285,150)
\Line(200,100)(270,30)
\GCirc(100,100){10}{1}
\GCirc(200,100){10}{1}
\GCirc(30,170){10}{1}
\GCirc(15,150){10}{1}
\GCirc(30,30){10}{1}
\GCirc(270,170){10}{1}
\GCirc(285,150){10}{1}
\GCirc(270,30){10}{1}
\DashCArc(100,100)(80,160,210){5}
\DashCArc(200,100)(80,-30,20){5}
%\Text(100,10)[]{Two centers with two legs each}
\end{picture}
\end{center}

\bea M^2 & = & \f{1}{a^2} \left( \ba{cccccccccc}
1 & 0 & \cdots & 0 & -1 & 0 & 0 & \cdots & 0 & 0 \\
0 & 1 & \cdots & 0 & -1 & 0 & 0 & \cdots & 0 & 0 \\
\cdots & \cdots & \cdots & \cdots & \cdots & \cdots & \cdots & \cdots
& \cdots & \cdots \\
0 & 0 & \cdots & 1 & -1 & 0 & 0 & \cdots & 0 & 0 \\
-1 & -1 & \cdots &-1 & N+1 & -1 & 0 & \cdots & 0 & 0 \\
0 & 0 & \cdots & 0 &  -1 & N+1 & -1 & \cdots & -1 & -1 \\
0 & 0 & \cdots & 0 & 0 & -1 & 1 & \cdots & 0 & 0 \\
\cdots & \cdots & \cdots & \cdots & \cdots & \cdots & \cdots & \cdots
& \cdots & \cdots \\
0 & 0 & \cdots & 0 & 0 &  -1 & 0 & \cdots & 1 & 0 \\
0 & 0 & \cdots & 0 & 0 & -1 & 0 & \cdots & 0 & 1 \ea \right) \eea

We can list the eigenvalues and the eigenstates for $\hat{M}^2 =
a^2 M^2$ up to normalization.
For large $N$ ( $N \gg 1$), the expression can be approximated as follows.

\bea \l = 0 & (1,1,\cdots,1,1,1,1,\cdots,1,1) \nn \\
\l = \f{2}{N} & (1,1,\cdots,1,1-\f{2}{N},-1+\f{2}{N},-1,\cdots, -1,-1) \nn \\
\l = 1 & (1,-1,\cdots,0,0,0,0,\cdots,0,0) \nn \\
& \cdots \nn \\
& (1,1,\cdots,-N+1,0,0,0,\cdots,0,0) \nn \\
&({\rm N-1}) \nn \\
\l = 1 & (0,0,\cdots,0,0,0,1,-1,\cdots,0) \nn \\
& \cdots \nn \\
& (0,0,\cdots,0,0,0,1,1,\cdots,-(N-1)) \nn \\
& ({\rm N-1}) \nn \\
\l = N+1 & (1,1,\cdots,1,-N,-N,1,\cdots,1,1) \nn \\
\l = N+3-\f{2}{N} &
(1,1,\cdots,1,-N-2+\f{2}{N},N+2-\f{2}{N},-1,\cdots,-1,-1) \nn
\eea

The presence of light modes $\l = \f{2}{N}$ is the most striking aspect
of two centers model. When there is a unique center,
the lightest excitation started from 1. Now it starts from $\f{2}{N}$
which is very light for $N \gg 1$.
Interpretation of the result is simple. If we disconnect the middle line
connecting two centers, we end up with two 'N-Octopus'
and each one has a zero mode. If we connect two centers with a new line,
it becomes a coupled system which mimics two ground state problem
in quantum mechanics. If there is a small mixing, the true ground state
is an even combination of two ground states and there is an excited state
which is an odd combination of the two ground states.
If the mixing vanishes, there are twofold degenerate ground state.
Here the middle line plays a role of the mixing between two states
and we get one zero mode (even combination of
each N-octopus zero mode) and one light mode 
(odd combination of each one).
\footnote{The author thanks R. Rattazzi for this simple interpretation.}

%%%%%%%%%%%%%%%%%%%%%%%%%%%%%%%%%%%%%%%%%%%%%%%%%%%%%%%%
\subsection{3 Legs with multiple sites}
%%%%%%%%%%%%%%%%%%%%%%%%%%%%%%%%%%%%%%%%%%%%%%%%%%%%%%%%

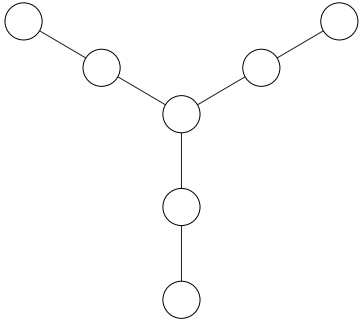
\SetScale{0.7}
\begin{center}
\begin{picture}(300,200)(100,50)
\Line(200,200)(115,250)
\Line(200,200)(285,250)
\Line(200,200)(200,100)
\GCirc(200,200){10}{1}
\GCirc(157,225){10}{1}
\GCirc(115,250){10}{1}
\GCirc(243,225){10}{1}
\GCirc(285,250){10}{1}
\GCirc(200,150){10}{1}
\GCirc(200,100){10}{1}
%\Text(150,80)[]{N-Octopus moose diagram}
\end{picture}
\end{center}

\bea M^2 & = & \f{1}{a^2} \left( \ba{ccccccc}
1 & -1 & 0 & 0 & 0 & 0 & 0 \\
-1 & 2 & 0 & 0 & 0 & 0 & -1 \\
0 & 0 & 1 & -1 & 0 & 0 & 0 \\
0 & 0 & -1 & 2 & 0 & 0 & -1 \\
0 & 0 & 0 & 0 & 1 & -1 & 0 \\
0 & 0 & 0 & 0 & -1 & 2 & -1 \\
0 & -1 & 0 & -1 & 0 & -1 & 3 \ea \right) \eea

We can list the eigenvalues and the eigenstates for $\hat{M}^2 =
a^2 M^2$ up to normalization.

\bea \l = 0 & (1,1,1,1,1,1,1) \nn \\
\l = \f{3-\sqrt{5}}{2} & (1,\f{\sqrt{5}-1}{2},-1,
\f{-\sqrt{5}+1}{2},0,0,0) \nn \\
& (1,\f{\sqrt{5}-1}{2},1,\f{\sqrt{5}-1}{2},
-2, -\sqrt{5}+1,0) \nn \\
\l = \f{3+\sqrt{5}}{2} & (1,\f{-\sqrt{5}-1}{2},-1,
\f{\sqrt{5}+1}{2},0,0,0) \nn \\
& (1,\f{-\sqrt{5}-1}{2},1,\f{-\sqrt{5}-1}{2},
-2,
\sqrt{5}+1,0) \nn \\
\l = 3-\sqrt{2} & (1,-2+\sqrt{2},1,-2+\sqrt{2},
1,-2+\sqrt{2},3-3\sqrt{2}) \nn \\
\l = 3+\sqrt{2} & (1,-2-\sqrt{2},1,-2-\sqrt{2},
1,-2-\sqrt{2},3+3\sqrt{2}) \nn \eea

The result shows that the addition of nodes provides more modes
which are heavier than the energy scale corresponding to 
the inverse of each leg. One clear thing is that there is no mode
whose scale is about $1/(6a)^2$.

%%%%%%%%%%%%%%%%%%%%%%%%%%%%%%%%%%%%%%%%%%%%%%%%%%%%%%%%
\subsection{3 Legs with multiple sites (different lengths)}
%%%%%%%%%%%%%%%%%%%%%%%%%%%%%%%%%%%%%%%%%%%%%%%%%%%%%%%%

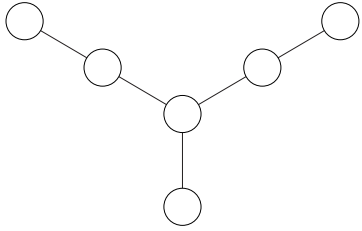
\SetScale{0.7}
\begin{center}
\begin{picture}(300,200)(100,100)
\Line(200,200)(115,250)
\Line(200,200)(285,250)
\Line(200,200)(200,150)
\GCirc(200,200){10}{1}
\GCirc(157,225){10}{1}
\GCirc(115,250){10}{1}
\GCirc(243,225){10}{1}
\GCirc(285,250){10}{1}
\GCirc(200,150){10}{1}
%\Text(150,80)[]{N-Octopus moose diagram}
\end{picture}
\end{center}

\bea M^2 & = & \f{1}{a^2} \left( \ba{cccccc}
1 & 0 & 0 & 0 & 0 & -1 \\
0 & 1 & -1 & 0 & 0 & 0 \\
0 & -1 & 2 & 0 & 0 & -1 \\
0 & 0 & 0 & 1 & -1 & 0 \\
0 & 0 & 0 & -1 & 2 & -1 \\
-1 & 0 & -1 & 0 & -1 & 3 \ea \right) \nn \eea

We can list the eigenvalues and the eigenstates for $\hat{M}^2 =
a^2 M^2$ up to normalization.

\bea \l = 0 & (1,1,1,1,1,1) \nn \\
\l = \f{3-\sqrt{5}}{2} & (0,1,\f{\sqrt{5}-1}{2},-1,
\f{-\sqrt{5}+1}{2},0) \nn \\
\l = \f{5-\sqrt{13}}{2} & (1,-\f{1}{2},\f{3-\sqrt{13}}{4},
-\f{1}{2},\f{3-\sqrt{13}}{4},\f{\sqrt{13}-3}{4}) \nn \\
\l = 2 & (1,1,-1,1,-1,-1) \nn \\
\l = \f{3+\sqrt{5}}{2} & (0,1,\f{-\sqrt{5}-1}{2},-1,
\f{\sqrt{5}+1}{2},0) \nn \\
\l = \f{5+\sqrt{13}}{2} & (1,-\f{1}{2},\f{3+\sqrt{13}}{4},
-\f{1}{2},\f{3+\sqrt{13}}{4},\f{-3-\sqrt{13}}{2}) \nn \eea

%%%%%%%%%%%%%%%%%%%%%%%%%%%%%%%%%%%%%%%%%%%%%%%%%%%%%%%%
\subsection{Large Extra Dimensions}
%%%%%%%%%%%%%%%%%%%%%%%%%%%%%%%%%%%%%%%%%%%%%%%%%%%%%%%%

Although we can not apply the deconstructed result
directly to gravity, the field theory analysis would give the same
result. Now the Kaluza-Klein states appear at very high scales.

Deviation of Newtonian potential can be understood in terms of
4 dimensional effective theory.
With massless graviton only, the potential between two test particles
with mass $m_1$ and $m_2$ separated by distance $r$ is
\bea
\f{V}{G_N m_1 m_2} & = & \f{1}{r}. 
\eea
If we consider 5 dimensional theory compactified on a circle
with radius $R$,
we have extra massive states with $M_n = \f{n}{R}$
for $n=1,2,\cdots$.
They also mediate gravitational interactions by Yukawa potentials
\bea
\f{\delta V}{G_N m_1 m_2} & = & \sum_{n=1}^{\infty}
\f{e^{-M_n r}}{r} \nn \\
& = & \f{e^{-\f{r}{R}}}{r(1-e^{-\f{r}{R}})}.
\eea
If $r \gg R$, $\f{\delta V}{V} \ll 1$ and we just have 4 dimensional
gravity.
However, if $r \ll R$, $\f{\delta V}{V} \gg 1$ and
$\f{\delta V}{G_N m_1 m_2} \simeq \f{R}{r^2}$ and
\bea
V & \simeq G^{(5)}_N \f{m_1 m_2}{r^2},
\eea
which produce 5 dimensional gravitational potential
($G^{(5)}_N = RG_N$, 5 dimensional Newton's constant).

We can do the same thing for higher dimensions
but now the exact summation formula is not available.
When $r \ll R$, we can approximate the summation with integrals
\bea
\f{\delta V}{G_N m_1 m_2} & = & \sum_{n_1,n_2,\cdots,n_{D-4}=1}^{\infty}
\f{e^{-M_n r}}{r} \nn \\
& = & \int_{n=1}^{\infty} dn C_{D-4} n^{D-5} \f{e^{- \f{nr}{R}}}{r} \nn \\
& = & C^{\prime}_{D-4} \f{R^{D-4}}{r^{D-3}} \nn
\eea
where $M_n = n/R$ with $n = \sqrt{n_1^2 + n_2^2 + \cdots n_{D-4}^2}$
for the isotropic compactification
($R_1 = R_2 = \cdots = R_{D_4} = R$).
$C_{D-4}$ is the solid angle of $D-4$ dimension.

Now let us consider the 'N-Octopus' configuration.
If we consider the setup in which $N$ equal length intervals
with size $\pi \rho$
attached at a single point (total length $= \pi R = \pi N \rho$),
the Kaluza-Klein spectrum comes as $N$ degenerate states
at $M_n = n/\rho$.
In this case Newtonian potential is modified by
\bea
\f{\delta V}{G_N m_1 m_2} & = & \sum_{n=1}^{\infty}
N \f{e^{-{M_n r}}}{r} \nn \\
& = & N \f{e^{-\f{r}{\rho}}}{r(1-e^{-\f{r}{\rho}})},
\eea
and when $r \ll \rho$, we have
\bea
\f{\delta V}{G_N m_1 m_2}
& = & \f{N \rho}{r^2} = \f{R}{r^2}.
\eea

Therefore, we can conclude that it just reproduces 5 dimensional gravity
when $r \ll \rho \ll R = N\rho$.
Note the relation between $R$ and $\rho$.
If $N \gg 1$, there is a huge difference between the scales
at which the gravity is modified and the scale that enters
in the modified potential.
The correction from massive gravitons become of order one
if $\f{\delta V}{V} \sim {\cal O}(1)$
and it is when the critical radius $r_c \simeq \r \log N$
which is not so much different from $\r$.
The scale entering in 5D potential is $R = N \r$
which is much larger distance scale than $\r$ or $\r \log N$.
In this way we can simple imagine 5D flat extra dimensional model
in which the fundamental scale is around TeV
while avoiding the phenomenological constrants from
the experiments.

We can choose $N$ large enough to make a single extra dimension scenario
be consistent with the current experimental bound.
For $r_c \simeq 0.1mm$ ($1/r_c \simeq 10^{-3}$ eV), 
if $1/\r = 10^{-1}$ or $10^{-2}$ eV
and $N = 10^{16}$ or $10^{17}$, we can explain the weak scale quantum gravity
with only single extra dimension.

On the other hand, the most stringent bound on the extra dimension
comes from supernovae and neutron stars. This bound is not applicable
if KK mass is heavier than 100 MeV.
Thus for $1/\r = 100$ MeV and $N = 10^{25}$, we start to see the fifth
dimension when $1/r_c \sim 1$ MeV and the gravity becomes strong
at TeV. Octopus configuration with large $N$ can avoid bounds on
large extra dimensions coming from light KK modes while
having TeV scale quantum gravity. The geometry considered here
postpone the appearance of KK modes till very short distance
(high energy) and all the modes appear at the same time
at very high energies.

%%%%%%%%%%%%%%%%%%%%%%%%%%%%%%%%%%%%%%%%%%%%%%%%%%%%%%%%
%\subsection{Gauge Theory on Extra Dimensions}
%%%%%%%%%%%%%%%%%%%%%%%%%%%%%%%%%%%%%%%%%%%%%%%%%%%%%%%%

%%%%%%%%%%%%%%%%%%%%%%%%%%%%%%%%%%%%%%%%%%%%%%%%%%%%%%%%
\subsection{4 fermi interactions}
%%%%%%%%%%%%%%%%%%%%%%%%%%%%%%%%%%%%%%%%%%%%%%%%%%%%%%%%

Unlike the usual case in which the first KK state appears at
$M_{KK} = 1/R$ and we get $1/M_{KK}^2$ after integrating out KK
states, here the KK states are extremely heavy, $M_{KK} = 1/\r =
N/R$. As there appear N such KK states, after integrating out KK
states, we get $1/M_{KK}^2 = 1/(NR^2)$ which is suppressed by $N$.
There would be many interesting phenomenology associated with it.

%%%%%%%%%%%%%%%%%%%%%%%%%%%%%%%%%%%%%%%%%%%%%%%%%%%%%%%%
\subsection{Warped Extra Dimension}
%%%%%%%%%%%%%%%%%%%%%%%%%%%%%%%%%%%%%%%%%%%%%%%%%%%%%%%%

It would be interesting to see what happens in the warped extra
dimensions.
We can analyze the spectrum of multi-throat configuration
in a similar way, but the result is not as interesting
as in flat space.
There is a single zero mode whose wave function is all over
the extra dimension. Then the excited states appear 
with wave functions localized near the throats
(especially when the curvature is large
which is distintively different from flat extra dimensions).
It is clearly seen in deconstruction setup
\cite{Falkowski:2002cm}\cite{Randall:2002qr}.
Gauge theory in a warped background has a nontrivial warp factor
in front of $\eta^{\mu \nu} F_{\mu 5} F_{\nu 5}$
and it can be deconstructed with a position dependent link VEV
$\langle \Phi_i \rangle = \langle \Phi_0 \rangle \e^i$ where
$\epsilon$ corresponds to $e^{-k/\Lambda}$ with $k$ the $AdS_5$ curvature
and $\Lambda$ the cutoff of the theory with $\e \ll 1$
for highly curved $AdS_5$
\cite{Falkowski:2002cm}.
The mass matrix for N sites is 
\bea
M^2 & = & \f{\e^2}{a^2} \left( 
\ba{ccccccc}
1 & -1 & 0 & 0 & 0 & \cdots & 0 \\
-1 & 1+\e^2 & -\e^2 & 0 & 0 & \cdots & 0 \\
0 & -\e^2 & \e^2 + \e^4 & -\e^4 & 0 & \cdots & 0 \\
\cdots & \cdots & \cdots & \cdots & \cdots & \cdots & \cdots \\
0 & \cdots & 0 & 0 & - \e^{2(N-3)} & \e^{2(N-3)} +\e^{2(N-2)} & -\e^{2(N-2)} \\
0 & \cdots & 0 & 0 & 0 & -\e^{2(N-2)} & \e^{2(N-2)} 
\ea \right)
\eea
The zero mode eigenstate is
\bea
A^{(0)}_\mu & = & \f{1}{\sqrt{N}} \sum_{i=1}^N A_{\mu,i}. 
\eea
For the excited states, the analysis is extremely simplified
when AdS is highly curved, $\e \ll 1$.
The higher mode eigenstates are
\bea
A^{(N-1)}_\mu & = & \f{1}{\sqrt{2}} ( A_{\mu,1} - A_{\mu,2}), \nn \\
A^{(N-2)}_\mu & = & \f{1}{\sqrt{6}} ( A_{\mu,1} + A_{\mu,2} - 2 A_{\mu,3}), 
\nn \\
\cdots \nn \\
A^{(1)}_\mu & = & \f{1}{\sqrt{(N-1)N}} 
( A_{\mu,1} + \cdots + A_{\mu,N-1} - (N-1) A_{\mu,N}), \nn 
\eea
where the coefficients are  determined up to ${\cal O}(\e^2)$. 
$A_\mu^{(N-j)}$ has the eigenvalue of order 
${\cal O}(\langle \Phi_0 \rangle \e^j)$
for $j=1,\cdots,N-1$
and the 5D interpretation is clear. For higher modes 
($m_n \sim \langle \Phi_0 \rangle$),
the wave function is localized near the UV brane.
The lightest mode is mostly localized near the IR brane.

The same analysis can be done for the multi-throat configuration
which has several IR branes and one UV brane with an Octopus shape.
For simplicity, let us consider two IR branes connected to the UV brane.
The mass matrix is then
\bea
M^2 & = & \f{\e^2}{a^2} \left( 
\ba{ccccccc}
\e^{2(N-2)} & \cdots & 0 & 0 & 0 & \cdots & 0 \\
\cdots & \cdots & \cdots & \cdots & \cdots & \cdots & \cdots \\
0 & \cdots & 1+\e^2 & -1 & 0 & \cdots & 0 \\
0 & \cdots & -1 & 2 & -1 & \cdots & 0 \\
0 & \cdots & 0 & -1 & 1+ \e^2 & \cdots & 0 \\
\cdots & \cdots & \cdots & \cdots & \cdots & \cdots & \cdots \\
0 & \cdots & 0 & 0 & 0 & \cdots & \e^{2(N-2)} 
\ea \right)
\eea
We can get the eigenstates from the simple UV-IR case.
We have 2N-1 sites and there are 2N-1 eigenstates.
The zero mode is flat along the extra dimension
which is the same as before.
The remaining 2N-2 modes are obtained simply by considering
even and odd combinations of two N-1 modes.
For instance, the lightest modes except the zero mode are
\bea
A^{(1+)}_\mu & = & \f{1}{\sqrt{2(N^2-N+3)}} 
( -(N-1) A_{\mu,1} + \cdots + A_{\mu,N-1} + 2A_{\mu,N} \nn \\
&&
+A_{\mu,N+1} + \cdots - (N-1) A_{\mu,2N-1}),\\ 
A^{(1-)}_\mu & = & \f{1}{\sqrt{2(N^2-N-1)}} 
( -(N-1) A_{\mu,1} + \cdots + A_{\mu,N-1} \nn \\
&&
-A_{\mu,N+1} + \cdots + (N-1) A_{\mu,2N-1}), 
\eea
up to ${\cal O}(\e^2)$,  
and the corresponding eigenvalues are degenerate (twofold degeneracy)
\bea
m_n & = & g\langle \Phi_0 \rangle \e^{(N-1)} \nn 
\eea
up to ${\cal O}(\e^2)$.  
More precisely the degeneracy is lifted by $1/N$ correction.
All the higher modes are similarly obtained and only for the heaviest one,
the eigenvalues are $m_n = g\langle \Phi_0 \rangle \e$ and 
$m_n = \sqrt{3} g\langle \Phi_0 \rangle \e$ up to 
${\cal O} (\e^2)$.

Therefore, the presence of the extra throat does not affect
the spectrum of lighter KK states.
Only when the KK mass is larger or comparable to the curvature scale,
the wave function connects different throats
and we get similar results as in flat extra dimensions.
This can be easily understood from AdS/CFT correspondence
\cite{Maldacena:1997re}\cite{Arkani-Hamed:2000ds}\cite{Rattazzi:2000hs}.
Each throat corresponds to a strongly coupled CFT
and each CFT has many resonances (KK modes).
The resonances in one CFT is nothing to do with
the ones in the other CFT. Thus KK spectrum in AdS
which corresponds to the resonances of CFT should not be affected
by the presence of other throats.

%%%%%%%%%%%%%%%%%%%%%%%%%%%%%%%%%%%%%%%%%%%%%%%%%%%%%%%%
\section{Field Theory Analysis}
%%%%%%%%%%%%%%%%%%%%%%%%%%%%%%%%%%%%%%%%%%%%%%%%%%%%%%%%

It is fairly simple to do the field theory analysis. As the analysis
is independent of Lorentz index, let us consider a massless scalar field
$\phi$ in 5 dimensions. The same result will be obtained for massless
vector fields, massless gravitons and massless fermions.

%%%%%%%%%%%%%%%%%%%%%%%%%%%%%%%%%%%%%%%%%%%%%%%%%%%%%%%%
\subsection{Octopus with N legs}
%%%%%%%%%%%%%%%%%%%%%%%%%%%%%%%%%%%%%%%%%%%%%%%%%%%%%%%%

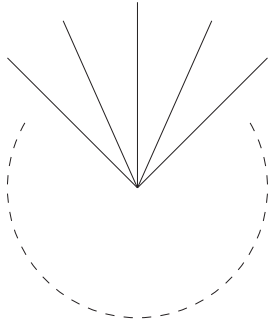
\SetScale{0.7}
\begin{center}
\begin{picture}(300,200)(100,50)
\Line(200,200)(200,300)
\Line(200,200)(160,290)
\Line(200,200)(240,290)
\Line(200,200)(130,270)
\Line(200,200)(270,270)
\DashCArc(200,200)(70,-210,30){5}
%\Text(150,80)[]{N-Octopus moose diagram}
\end{picture}
\end{center}

First of all, we consider a joint of N intervals at a single point.
The figure shows a schematic configuration and dots represent omitted
N-5 intervals. The figure just shows the extra dimension
and the relative angle between two intervals or the ordering of
different intervals do not have any physical meaning in the configuration
as there is no space at all beyond the extra dimension denoted by
lines in the figure.
The lagrangian for a massless scalar field is
\bea
{\cal L} & = & \int d^4x \left( \int_0^{2\pi \r} dx_5^{(1)}
+ \int_0^{2\pi \r} dx_5^{(2)} + \cdots 
+\int_0^{2\pi \r} dx_5^{(N)} \right) 
\left[ \f{1}{2} \partial_M \phi (x,x_5) \partial^M \phi (x,x_5) \right], 
\eea
where $M=0,1,2,3,5$ is 5 dimensional Lorentz index.
For the octopus of N legs with Neumann boundary conditions at N ends
of the legs (for simplicity, we assume all the legs are equal in length,
$\pi \r$ ),
\bea
\f{\partial}{\partial x_5^{(i)}} \phi^{(i)} (x_\mu, x_5^{(i)} =0) & = & 0, 
\eea
at $x_5^{(i)}=0$ with $i=1,\cdots,N$.
We restrict our analysis to the case when there is no localized term
at the junction.
The remaining boundary conditions are i) the wave function should be
continuous (as we do not have any extra terms located at special points)
and ii) the derivatives should cancel.
The first and the second conditions are
\bea
&& \phi^{(i)} (x_\mu, x_5^{(i)}=\pi \r) = \phi^{(j)} (x_\mu, x_5^{(j)}=\pi \r),
\\
&& \sum_{i=1}^N
\f{\partial}{\partial x_5^{(i)}} \phi^{(i)} (x_\mu, x_5^{(i)} =\pi \r) = 0. 
\eea
Here we introduce coordinates $x_5^{(i)}$ $(i=1,\cdots,N)$ which runs
from 0 (the end of the $i$th leg) to $\pi \r$ (the center/junction).
We are ready to find the spectrum.
Let
\bea
\phi^{(i)} (x_\mu, x_5^{(i)}) & = & \sum_n A^{(i)} \phi_n^{(i)} \cos
(k_n x_5^{(i)}).
\eea
The boundary condition at the ends of the legs are satisfied.
The remaining boundary conditions are $N-1$ conditions for the wave
functions at the junction and one condition for the cancellation of derivatives
at the junction.

As the junction is located at $x_5^{(i)}=\pi \r$ for all $i$
(equal distance away from the ends), the boundary condition is
\bea
A^{(i)} \cos (k_n^{(i)} \pi \r) & = & A^{(j)} \cos (k_n^{(j)} \pi \r).
\eea
which can be satisfied either for i) $k_n^{(i)} \pi \r = (n^{(i)}+\f{1}{2}) \pi$
or ii) $A^{(i)} = A^{(j)}$ for all $i \neq j$.
For i), the final boundary condition is
\bea
\sum_i A^{(i)} (-1)^{n^{(i)}+1} & = & 0. 
\eea
For ii), the condition is
\bea
(\sum_i A^{(i)} ) \sin (k_n^{(i)} \pi \r)
& = & N A^{(1)} \sin (k_n^{(i)} \pi \r), 
\eea
and it can be satisfied only when $k_n^{(i)} \pi \r = n^{(i)} \pi$
since $A^{(1)} \neq 0$.

Now all the eigenvalues are determined.
Let us consider how many degenerate states are there for each $k_n$.
For i), we have $N-1$ independent solutions which can be written
in terms of a N dimensional vector $v$
\bea
v & = & (A^{(1)},A^{(2)},\cdots,A^{(N)}). 
\eea
as
\bea
v & = & \f{1}{\sqrt{2}} \sqrt{\f{2}{\pi \r}} (1,-1,0,\cdots,0) \nn \\
 & &  \f{1}{\sqrt{6}} \sqrt{\f{2}{\pi \r}} (1,1,-2,\cdots,0) \nn \\
 & & \cdots \nn \\
 & & \f{1}{\sqrt{(N-1)N}} \sqrt{\f{2}{\pi \r}} (1,1,1,\cdots,-(N-1)) \nn
\eea

For ii), all the coefficients are determined and there is a single state.
\bea
v & = & \f{1}{\sqrt{N}} \sqrt{\f{2}{2^{\delta_{n,0}} \pi \r}}
(1,1,1,\cdots,1) \nn
\eea

We should be careful here.
For ii), we can imagine a wave function which is connected
with different $n^{(i)}$s at differnt $x_5^{(i)}$.
As we know that there is a zero mode with a flat potential,
we can check whether the arbitrary $n^{(i)}$ can yield
the orthogonality condition.
For $n^{(i)} \neq n^{(j)}$, the wave functions are orthogonal
at the $i$th leg.
The lightest mode (except the zero mode) should not include
$n^{(i)}=0$ as they will generate a nonzero positive contribution
when we consider orthogonality condition with the zero mode.
$n^{(i)} \ge 1$ is required from the consideration
and the lightest mode is $n^{(i)}=1$ for all $i$.
Similar reasoning gives $n^{(i)}=2$ and higher
and we can simply replace $n^{(i)}=n$.

Now the spectrum is alternating. We have a single mode at $M_n = n/\r$
and $N-1$ modes in between $n$th and $n+1$th mode ($M_n = (n+\f{1}{2})/\r)$).

Asymmetry between the degeneracy of $n/\r$ and $(n+\f{1}{2})/\r$ modes
can be understood as follows. We put Neumann boundary conditions
at the ends of the legs and thus the states with Dirichlet boundary conditions
are projected out.

If we impose Dirichlet boundary condition at the ends of the legs,
we would encounter the opposite case. There is no zero mode
and a single mode at $M_n = (n+\f{1}{2})/\r$ and $N-1$ modes
at $M_n = (n+1)/\r$ with $n \ge 0$.

%%%%%%%%%%%%%%%%%%%%%%%%%%%%%%%%%%%%%%%%%%%%%%%%%%%%%%%%
\subsection{Flower with N leaves}
%%%%%%%%%%%%%%%%%%%%%%%%%%%%%%%%%%%%%%%%%%%%%%%%%%%%%%%%

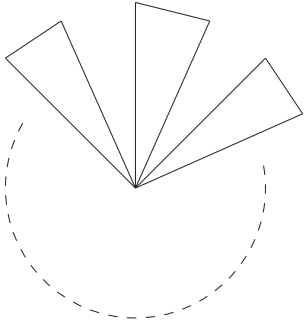
\SetScale{0.7}
\begin{center}
\begin{picture}(300,200)(100,50)
\Line(200,200)(200,300)
\Line(200,200)(160,290)
\Line(200,200)(240,290)
\Line(200,200)(130,270)
\Curve{(130,270)(160,290)}
\Curve{(200,300)(240,290)}
\Curve{(270,270)(290,240)}
\Line(200,200)(270,270)
\Line(200,200)(290,240)
\DashCArc(200,200)(70,-210,10){5}
%\Text(150,80)[]{N-Octopus moose diagram}
\end{picture}
\end{center}

To see the picture clearly, let us consider a flower configuration
where N rings are attached at the same point (center).
Each ring has a circumference $2\pi \r$. We can do the similar analysis.
Now $x_5^{(i)}$ is from $0$ to $2\pi \r$ and
\bea
\phi^{(i)} (x_\mu, x_5^{(i)}) & = & \sum_n (
A^{(i)} \phi_n^{(i)} \cos (k_n x_5^{(i)})
+ B^{(i)} \phi_n^{(i)} \sin (k_n x_5^{(i)}) ).
\eea

For each ring(leaf), the boundary condition corresponding to
the end points of Octopus is
\bea
\phi^{(i)} (x_5^{(i)} + 2\pi \r) & = &  \phi^{(i)} (x_5^{(i)} ),
\eea
and it determines $k_n^{(i)} 2\pi \r = 2\pi n^{(i)}$
and $k_n^{(i)} = n^{(i)}/\r$.
The remaining boundary condition at the center is the same.
If we assign the center to be $x_5^{(i)} = \pi \r$, the first $N-1$ boundary
condition requires
\bea
A^{(i)} (-1)^{n^{(i)}+1} & = &  A^{(j)} (-1)^{n^{(j)}+1}. 
\eea

The special limit is when all $A^{(i)} = 0$.
The second boundary condition is automatically satisfied.
For each $\phi^{(i)}$, there are incoming and outgoing derivatives
which cancel with each other. Therefore, for each $n^{(i)}$,
we can have $N+1$ independent solutions except when $n^{(i)}=0$.
For $n^{(i)}=0$ for all $i$s, we have the usual zero mode.
\bea
v & = & \f{1}{\sqrt{N}} \f{1}{\sqrt{2\pi \r}} (1,1,1,\cdots,1) \nn \\
w & = & (0,0,0,\cdots,0) \nn
\eea

Note that you do not need to have the same $k_n^{(i)}$ for different $i$s.
The lightest mode appears when all $k_n^{(i)}=1$.
There are $N+1$ such states which are degenerate with $k_n = 1/\r$.
One is
\bea
v & = & \f{1}{\sqrt{N}} \f{1}{\sqrt{\pi \r}} (1,1,1,\cdots,1) \nn \\
w & = & (0,0,0,\cdots,0) \nn
\eea
and the other $N$ states are
\bea
v & = & (0,0,0,\cdots,0) \nn \\
w & = & \f{1}{\sqrt{\pi \r}} (1,0,0,\cdots,0) \nn \\
  & & \f{1}{\sqrt{\pi \r}} (0,1,0,\cdots,0) \nn \\
  & & \cdots \\
  & & \f{1}{\sqrt{\pi \r}} (0,0,0,\cdots,1) \nn
\eea

For the latter case, it can be thought that
the modes will be lighter than $n/\r$
as there is only one ring that gives Kaluza-Klein mass.
However, there is no wave function outside of the ring
and the result is the same as the case with a single ring with a radius $\r$.
You can see that there are $N+1$ states at each $n/\r$
except $n=0$ (a single zero mode).

%%%%%%%%%%%%%%%%%%%%%%%%%%%%%%%%%%%%%%%%%%%%%%%%%%%%%%%%
\subsection{Caterpillar}
%%%%%%%%%%%%%%%%%%%%%%%%%%%%%%%%%%%%%%%%%%%%%%%%%%%%%%%%

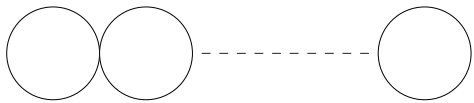
\SetScale{0.7}
\begin{center}
\begin{picture}(300,100)(0,0)
\GCirc(50,70){25}{1}
\GCirc(100,70){25}{1}
\GCirc(250,70){25}{1}
\DashLine(130,70)(220,70){5}
\end{picture}
\end{center}

Finally let us consider a ring that is attached with each other
but the ring intersects only with two nearest neighbor rings
(except the edge ring which intersects with only one ring).
It would be a sequence of shape 8 and let us call it 'caterpillar'.
From the boundary conditions
\bea
\phi^{(i)} (x_5^{(i)} + 2\pi \r) & = &  \phi^{(i)} (x_5^{(i)} ), 
\eea
we can determine $k_n^{(i)} = n^{(i)}/\r$.
When $A^{(1)} \neq 0$,
the wave function is continuous if
\bea
A^{(i)} (-1)^{n^{(i)}} & = & A^{(i+1)}. 
\eea

There is no condition for $B^{(i)}$ as the derivatives cancel
within the same ring.
The situation is the same as in the flower configuration.
The first one for $n=0$ is
\bea
v & = & \f{1}{\sqrt{N}} \f{1}{\sqrt{2 \pi \r}} (1,1,1,\cdots,1) \nn \\
w & = & (0,0,0,\cdots,0) \nn
\eea
and for $n^{(i)} \neq 0$,
\bea
v & = & \f{1}{\sqrt{N}} \f{1}{\sqrt{\pi \r}}
(1,(-1)^{n^{(i)}},1,\cdots,(-1)^{(n^{(i)} N)} ) \nn \\
w & = & (0,0,0,\cdots,0) \nn
\eea
and the other $N$ states are
\bea
v & = & (0,0,0,\cdots,0) \nn \\
w & = & \f{1}{\sqrt{\pi \r}} (1,0,0,\cdots,0) \nn \\
  & & \f{1}{\sqrt{\pi \r}} (0,1,0,\cdots,0) \nn \\
  & & \cdots \\
  & & \f{1}{\sqrt{\pi \r}} (0,0,0,\cdots,1) \nn
\eea
There are totally $N+1$ degenerate states for each $k_n = n/\r$.

However, there appears much lighter states in this case.
Suppose $N= 2k +1$. Then we can imagine a configuration in which
$k_n^{(i)} = 0$ for all $i$s except $i = k+1$
and $k_n^{(k+1)} = 1/\r$. The wave function is
\bea
v & = & \f{1}{\sqrt{N}} \f{1}{\sqrt{2^{\delta_{n^{(i)},0}} \pi \r}}
(1,1,1,\cdots,1,-1,\cdots,-1,-1,-1) \nn \\
w & = & (0,0,0,\cdots,0) \nn
\eea
Now as there is an $\f{1}{\sqrt{N}}$ volume suppression
in the wave function and the mode is still orthogonal to the zero mode.
The contributions of the first $k$ rings cancel the ones of the last
$k$ rings and $k+1$ ring wave function is orthogonal
since $n^{(k)}=0$ for the zero mode
and $n^{(k)}=1$ for the mode considered here.
The Kaluza-Klein mass only comes from a single ring
and we get $k^{4D} = 1/(\sqrt{N} \r)$ rather than $1/\r$.
Here the configuration is uniquely determined since
the change of the wave function ($n^{(i)} \neq 0$)
should be located in the middle to balance the wave function
such that it can be orthogonal to the zero mode.

If we consider $n^{(i)}=1$ for two $i$s,
we can not make the wave function to be orthogonal to the zero mode
if $N= 2k+1$.
Instead, we can consider $N=2k$.
As there are $2k-2$ rings with $n^{(i)}=0$,
they should be evenly divided into positive and negative amplitudes.
It is possible when the first nonzero $n^{(i)}$
and the second nonzero $n^{(i)}$ has a separation of $k-1$.
There are $k$ such possibilities.
Although it would be interesting to study the spectrum of these cases
in detail, we will not pursue it here.

You can see the huge difference between the flower and the caterpillar
configurations. There is no constraint from the derivative matching
and the momentum in one ring can be different from the one in the other ring
in principle. As a consequence the lightest mode start to appear
at $1/(\sqrt{N} \r)$ although the actual configuration is not a homogeneous
variation along $2\pi \r$ but a rapid variation only at a local region.
This is in accord with the deconstruction result of two centered Octopus.

We stress here that it is the presence of a junction from which
all the subsegments or rings are connected and they make it
possible to raise the scale of the Kaluza-Klein excitations.

%%%%%%%%%%%%%%%%%%%%%%%%%%%%%%%%%%%%%%%%%%%%%%%%%%%%%%%%
\section{Conclusion}
%%%%%%%%%%%%%%%%%%%%%%%%%%%%%%%%%%%%%%%%%%%%%%%%%%%%%%%%

In this paper we have shown that the spectrum of Kaluza-Klein particles
can be rich and interesting even with a single extra dimension.
Depending on how the extra dimension is connected with each other,
the KK spectrum appears entirely differently.
The most interesting aspect is that we can defer the appearance
of the lightest KK modes as high as we want.
This is impossible with a simple circle compactification
or an orbifolding of it.
With a single extra dimension the lightest KK mode is directly
linked to the size of the extra dimension (1/R)
if the extra dimension is a simple circle or an interval.
Several examples considered in this paper shows that
the relation no longer holds if several extra dimension is connected
with a common point, so called 'junction' or 'center'.
This enables us to have TeV scale quantum gravity with
a single extra dimension and a lightest KK mass of 100 MeV.

The fact that this new setup is just 5 dimensional spacetime
is important. 
This opens an entirely new era for figuring out
what would be the shape of the extra dimension relevant to the real world.
Before going to higher dimensional theory,
we can study a lot of examples with a single extra dimension.
Orbifold GUT in 6D \cite{Asaka:2001eh}\cite{Hall:2001xr}
\cite{Dermisek:2001hp}\cite{Kim:2002im}\cite{Kim:2004vk}
has more freedom over 5D model since we can use two
orbifolding parity and two Wilson lines.
However, now with the setup considered here, we can do the similar thing
in 5D by attaching two or three intervals.
Model building should be seriously done with these new setups.
It would be possible to build a simple model in 5D.
The configuration might be regarded as ad hoc.
However, it can be understood as an effective description 
of the underlying theory
and most of important physics questions can be addressed
without relying on what the exact underlying theory is.

Although we studied the single extra dimension only, 
the 'center' or the 'junction' can play the same role
when two or more spatial extra dimensions are attached.
Also the 'center' or the 'junction' can be generalized to arbitrary 
higher dimensions.
Furthermore, we have not introduced any local kinetic terms
or mass terms here but we can study the general cases
in which the 'center' has a special interactions.
Boundary conditions at the leg also can be generalized.
We leave many detailed example studies for the future work.
Phenomenological constraints on the extra dimension also should be restated
after considering several variations of the simple compactification.

%%%%%%%%%%%%%%%%%%%%%%%%%%%%%%%%%%%%%%%%%%%%%%%%%%%%%%%%
\section*{Acknowledgement}
%%%%%%%%%%%%%%%%%%%%%%%%%%%%%%%%%%%%%%%%%%%%%%%%%%%%%%%%

The author thanks N. Arkani-Hamed, G. Giudice, N. Kaloper, J. March-Russell
and R. Rattazzi for discussions and CERN for their hospitality during the visit.
This work is supported by the ABRL Grant No. R14-2003-012-01001-0,
the BK21 program of Ministry of Education, Korea
and the SRC Program of the KOSEF
through the Center for Quantum Spacetime
of Sogang University with grant number R11-2005-021.

%%%%%%%%%%%%%%%%%%%%%%%%%%%%%%%%%%%%%%%%%%%%%%%%%%%%%%%%

%%%%%%%%%%%%%%%%%%%%%%%%%%%%%%%%%%%%%%%%%%%%%%%%%%%%%%%%%

%%%%%%%%%%%%%%%%%%%%%%%%%%%%%%%%%%%%%%%%%%%%%%%%%%%%%%%%%
\end{document}